\documentclass[aip, pof,
 amsmath,
 amssymb,
reprint,
showpacs,
showkeys,
]{revtex4-1}
\usepackage{graphicx}
\pdfoutput=1
\usepackage{dcolumn}
\usepackage{bm}

\begin{document}
\title{Dynamics of unconfined spherical flames}
\author{L. Leblanc}
\author{M. Manoubi}
\author{K. Dennis}
\affiliation{Department of Mechanical Engineering, University of Ottawa, Canada}
\author{Z. Liang}
\affiliation{Atomic Energy Canada Limited, Chalk River, Canada}
\author{M. I. Radulescu}
\email{matei@uottawa.ca}
\affiliation{Department of Mechanical Engineering, University of Ottawa, Canada}
\date{\today}

\begin{abstract}
Using the soap bubble technique, we visualize the dynamics of unconfined hydrogen-air flames using high speed schlieren video.  We show that for sufficiently weak mixtures, i.e., low flame speeds, buoyancy effects become important.  Flame balls of a critical dimension begin to rise.  The experiments are found in very good agreement with the scaling laws proposed by Zingale and Dursi. We report the results in a fluid dynamics video. 
\end{abstract}

\maketitle
\section{The soap bubble technique for unconfined flames}
In accidental releases of combustible gases in unconfined spaces, deflagrations occur at nearly constant pressure. Constant pressure conditions are usually difficult to establish in the laboratory, as any confinement leads to a pressure increase during the flame propagation.  The soap bubble technique \cite{Stevens1926} permits to maintain a nearly constant pressure and simulate unconfined conditions.  The  bubble can be filled with a reactive mixture, which is then ignited. As the film cannot sustain large pressure differences across it, it thus acts as an ideal contact discontinuity separating the gases inside the bubble from the gases outside. \\
The present study addresses the dynamics of such weakly confined deflagrations.  We study a mixture of hydrogen and air using high speed schlieren visualization.  The accompanying video demonstrates the two regimes of propagation, briefly described below. 
\section{Strong mixtures}
\begin{figure}
\begin{center}
 \includegraphics[width=0.7\columnwidth]{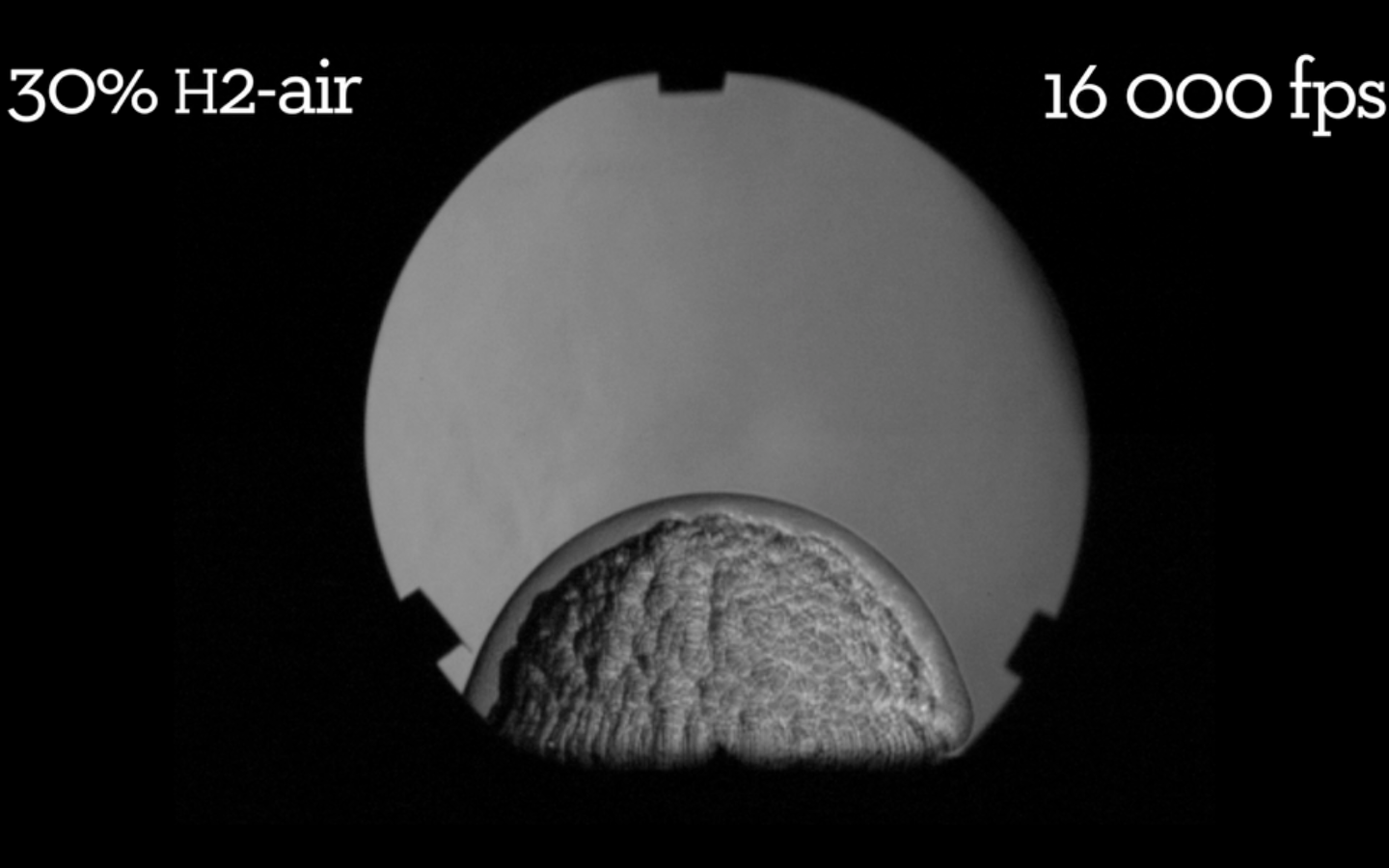}
\caption{A hemi-spherical flame in a $30\%H_2 - Air$ mixture contained in a soap bubble; for reference, the field of view of is 30 cm in diameter.} 
\label{fig:fig1}
\end{center}
\end{figure}

For near stoichiometric mixtures, the flame takes on a spherically symmetric structure, as shown in accompanying video. A select frame illustrating the flame structure before it encounters the soap bubble is shown in Figure \ref{fig:fig1}. The instabilities developed on the flame structure are also clearly discernable; these are well documented in the literature \cite{Jomaas2007}. 

\section{Weak mixtures} 
\begin{figure}
\begin{center}
 \includegraphics[width=0.7\columnwidth]{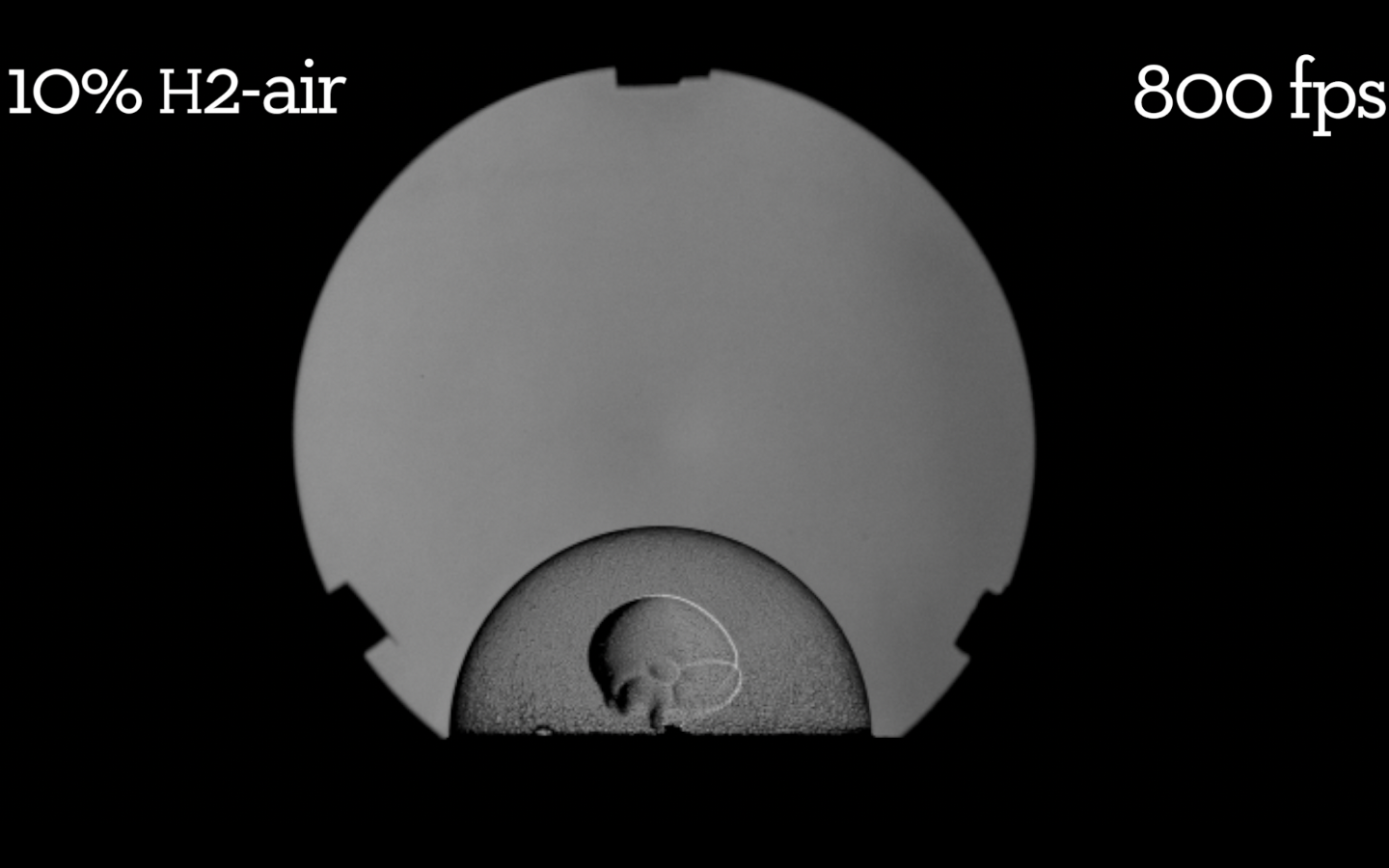}
 \includegraphics[width=0.7\columnwidth]{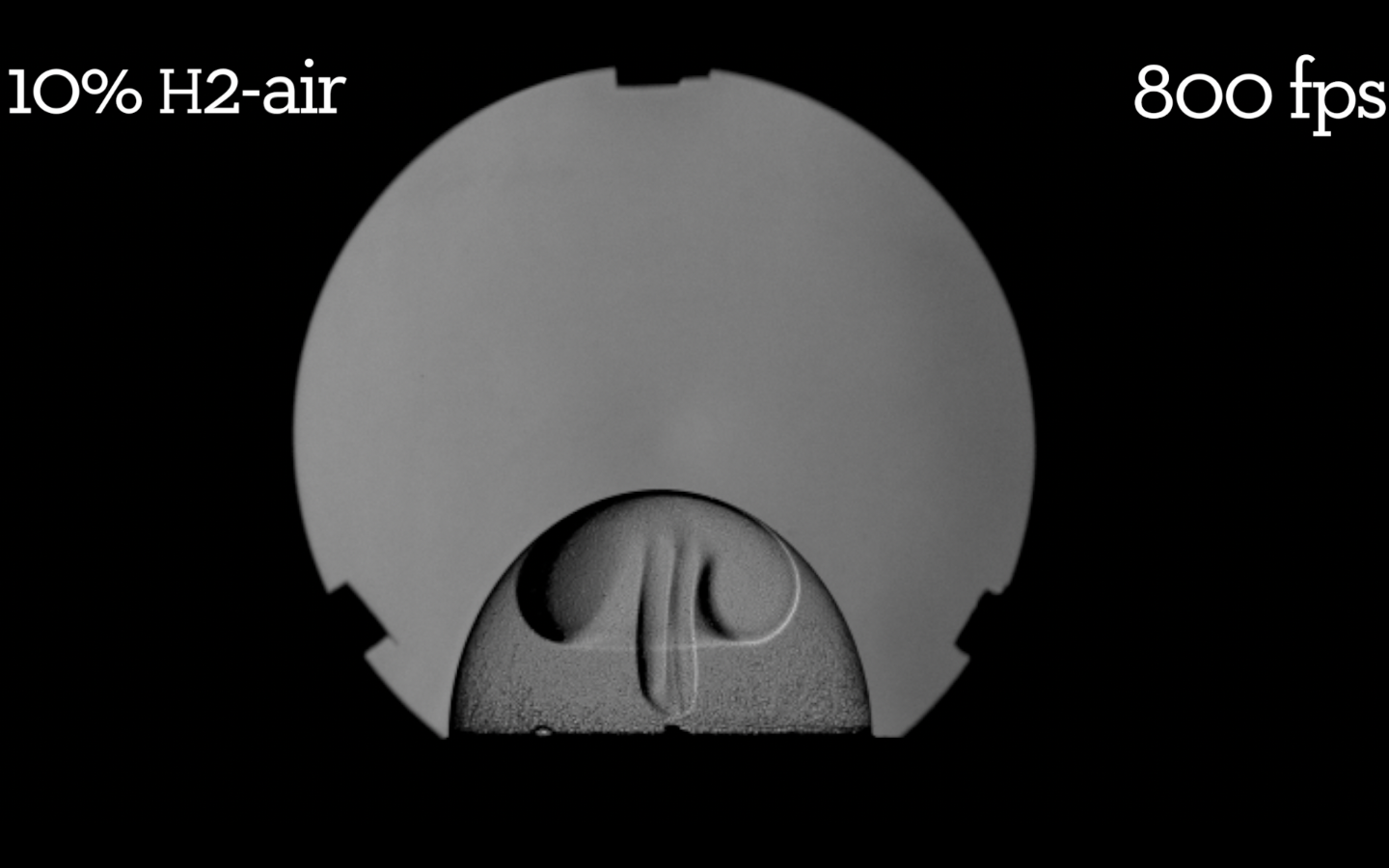}
\caption{A hemi-spherical flame in a $10\%H_2 - Air$ mixture contained in a soap bubble; for reference, the field of view of is 30 cm in diameter.} 
\label{fig:fig2}
\end{center}
\end{figure}

For weak mixtures, i.e., mixtures characterized by low flame speeds, we find that the flames remain spherical initially, but begin to rise. Figure \ref{fig:fig2} shows two such frames. This phenomenon is due to the gravitational acceleration (buoyancy forces) preferentially accelerating the light combustion products contained inside the flame ball.  Following the initial spherical regime, the flame develops a mushroom shape. The video demonstrates how large portions of unreacted material are left behind.  If ignition is effected a second time, a new flame kernel develops, which then starts racing through the first.

\section{Transition between the two regimes}
The two regimes of burning illustrated above are differentiated by the time scales associated with the flame motion via diffusive effects (i.e. flame propagation) and the time scale associated with the motion of the bubble due to buoyancy forces.  Following a similar treatment as Zingale and Dursi \cite{Zingale&Dursi2007}, these scales can be simply approximated as follows.\\
	Consider a reactive gas, characterized by a given laminar flame speed $S$ and expansion ratio $\rho_b / \rho_u $, where the subscripts \textit{b} and \textit{u} refer respectively to the burned and unburned gases. Continuity across the flame requires that the speed of the flame (with respect to the stationary burned gases) is $V_{burn}=(\rho_u/\rho_s)S$.  The time scale for the flame to grow to a characteristic dimension $R$ is thus
\begin{align}
t_{burn}=\frac{R}{V_{burn}}=\frac{R}{S}\frac{\rho_b}{\rho_u}
\end{align}
\\
The time scale for buoyancy forces to displace the flame bubble the same characteristic distance $R$ can be obtained from simple physical arguments.  The upwards motion of the bubble is governed by the competition between the buoyancy force, $4/3 \pi R^3 (\rho_u - \rho_b) g$ and the drag force.  Since drag for such rising bubbles is due mainly to pressure drag in inviscid flow (viscous effects are negligible), the drag is simply proportional to the characteristic dynamic pressure $1/2 \rho_u V_{rise}^2$ and the projected surface area of the flame bubble,  $\pi R^2$.  Equating the two forces, one obtains the characteristic rise speed of the bubble; adopting the numerical factor determined by Davies and Taylor \cite{Davies&Taylor1950} by appropriately accounting for the pressure distribution, we obtain
\begin{align}
V_{rise}=2/3 \sqrt{Rg \left(1- \frac{\rho_b}{\rho_u} \right)}
\end{align}
yielding a characteristic time scale for buoyancy effects of $t_{rise}=R/V_{rise}$.\\
The ratio of the characteristic time scales of burning and buoyancy yields
 \begin{align}
\theta = \frac{t_{burn}}{t_{rise}}=\frac{2}{3} S \frac{\rho_b}{\rho_u} \sqrt{R g \left(1- \frac{\rho_b}{\rho_u} \right)}
\end{align}
We thus expect the weak flame regime when $\theta >> 1$ and the strong regime when $\theta << 1$.  Alternatively, one can define a critical flame radius $R_{switch}$ at which the two time scales are equal \cite{Zingale&Dursi2007}.   Setting $\theta=1$, we obtain 
\begin{align}
R_{switch}= \frac{9}{4} \frac{S^2}{g} \left(\frac{\rho_b}{\rho_u}\right)^{-2} \left(1- \frac{\rho_b}{\rho_u} \right)^{-1}
\end{align}
As the flame grows from a small radius, it will not be influenced by buoyant forces until $R \simeq R_{switch}$.\\
	We can compare the above criterion with the two experiments shown in the accompanying video.  For a $30 \% H_2$ flame, the flame speed is approximately 2.5 m/s, and the expansion ratio is ${\rho_b}/{\rho_u}=0.14$.  This yields a critical radius of  85m.  Clearly, the flame will not be dominated by buoyancy effects on the length scales of the experiment ($10^{-1} m$).  For the $10 \% H_2$ flame, the flame speed is approximately 0.1 m/s, and the expansion ratio is ${\rho_b}/{\rho_u}=0.3$.  This yields a critical radius of 4 cm.  This is perfectly compatible with our experiments. We observe a spherical bubble detaching from the bottom wall when the flame radius reaches a few centimeters, as illustrated in Fig. \ref{fig:fig2}.
\bibliography{references}

\end{document}